\documentclass[preprint, showpacs,preprintnumbers,amsmath,amssymb,nofootinbib]{revtex4}

 \usepackage{epsf}
 \usepackage{graphicx}

\begin{document}
\newcommand{\be}[1]{\begin{equation}\label{#1}}
 \newcommand{\ee}{\end{equation}}
 \newcommand{\bea}{\begin{eqnarray}}
 \newcommand{\eea}{\end{eqnarray}}
 \def\disp{\displaystyle}

 \def\gsim{ \lower .75ex \hbox{$\sim$} \llap{\raise .27ex \hbox{$>$}} }
 \def\lsim{ \lower .75ex \hbox{$\sim$} \llap{\raise .27ex \hbox{$<$}} }

\title{\Large \bf The growth of linear perturbations in the DGP model}
\author{Xiangyun Fu$^{1}$, Puxun Wu$^{1, 2, 3}$ and Hongwei Yu$^{1, 2}$\footnote{Corresponding author}}

\address{$^1$Department of Physics and Institute of  Physics, Hunan
Normal University, Changsha, Hunan 410081, China
\\
$^2$Kavli Institute for Theoretical Physics China, CAS, Beijing
100190, China
\\
$^3$Department of Physics and Tsinghua Center for Astrophysics,
Tsinghua University, Beijing 100084, China }

 \begin{abstract}
We study the linear growth of matter perturbations in the DGP
model with the growth index $\gamma$ as a function of redshift. At
the linear approximation:
$\gamma(z)\approx\gamma_0+\gamma_0^\prime z $,  we find that, for
$0.2\leq\Omega_{m,0}\leq0.35$, $\gamma_0$ takes the value from
$0.658$ to $0.671$,   and $\gamma_0^\prime$ ranges from $0.035$ to
$0.042$. With three low redshift observational data of the growth
factor, we obtain the observational constraints on $\gamma_0$ and
$\gamma_0'$ for the $\Lambda CDM$ and DGP models and find that the
observations favor the $\Lambda CDM$ model but at the $1\sigma$
confidence level both the $\Lambda CDM$ and DGP models are
consistent with the observations.
\end{abstract}


 \maketitle
 \renewcommand{\baselinestretch}{1.5}

\section{Introduction}\label{sec1}
Various observations show that our universe is undergoing an
accelerating expansion~\cite{Sne, CMB, SDSS} and many  models have
been proposed to explain this mysterious phenomenon. There are
basically two main classes of models.  One is dark energy which
yields sufficient negative pressure to induce a late-time
accelerated expansion; the other is  the modified gravity, such as
the scalar-tensor theory~\cite{scalartensor}, the $f(R)$
theory~\cite{frmodel} and the Dvali-Gabadadze-Porrati (DGP)
braneworld scenarios~\cite{r32,r33}, et al. However, these models
may predict the same late time accelerated cosmological expansion,
although they are quite different physically. So an important task
is to discriminate one from another. Recently, some attempts have
been made~\cite{r1,r12,r13,r14,r15,r16,r17} in this regard. An
interesting approach is to differentiate the dark energy and the
modified gravity with the growth function
$\delta(z)\equiv\delta\rho_m/\rho_m$ of the linear matter density
contrast as a function of redshift $z$. While different models give
the same late time expansion,  they may produce different growth of
matter perturbations~\cite{aastarobinsky}.

To the linear order of perturbation,  the matter density
perturbation $\delta=\delta\rho_m/\rho_m$ satisfies the following
equation~\cite{wangli} at the large scales
\begin{equation}
\label{denpert} \ddot{\delta}+2H\dot{\delta}-4\pi
G_{eff}\,\rho_m\delta=0,
\end{equation}
where  $G_{eff}$ is the effective Newton's constant and the dot
denotes the derivative with respect to time $t$. In general
relativity,  $G_{eff}= G_N$ where $G_N$ is the Newton's constant.
Defining the growth factor $f\equiv d\ln\delta/d\ln a$, one can
obtain
\begin{equation}
\label{grwthfeq1} {d\; f\over d\ln
a}+f^2+\bigg(\frac{\dot{H}}{H^2}+2
\bigg)f=\frac{3}{2}\frac{G_{eff}}{G_{N}}\Omega_m,
\end{equation}
where $\Omega_m$ is the fractional energy density of matter. In
general, analytical solutions to Eq. (\ref{grwthfeq1}) are hard to
find, and we need to resort to numerical methods. It has been
known for many years that there is a good approximation to the
growth factor $f$,  which is given by~\cite{jnfry}
 \begin{equation}
 \label{fommegam}
 f\equiv\frac{d\ln\delta}{d\ln a}\simeq\Omega_m(z)^\gamma,
 \end{equation}
where $\gamma$ is the growth index and  is taken as a constant.
This parameterized approach has been studied in some works
recently, see
e.g.~\cite{r18,r19,r20,r22,r28,r29,r30,r31,weihao,bboisseau,gongyungui}.
For example, substituting the above equation into
Eq.~(\ref{grwthfeq1}) and then expanding around $\Omega_m=1$ (a
good approximation at the high redshift), one can obtain
$\gamma_\infty\simeq 0.5454$~\cite{r18,r19} for  the  $\Lambda
CDM$ model and $\gamma_\infty
\simeq11/16\approx0.6875$~\cite{r18,weihao} for the flat DGP
model. Therefore, in principle, one can distinguish the dark
energy model from the modified gravity model with observational
data on the growth factor. However, taking the index $\gamma$ as a
constant is only an approximation although it is a very good one
in certain circumstances. More generically, one should  rewrite
Eq.~(\ref{fommegam}) as
\begin{equation}
 \label{fommegam1}
 f\equiv\frac{d\ln\delta}{d\ln a}=\Omega_m(z)^{\gamma(z)}\;.
 \end{equation}
Defining a new quantity $\gamma^\prime\equiv{d \gamma(z)\over
 dz}$,  we can  expand $\gamma$ at the low redshift, as
 follows
\be{gammaz}
 \gamma(z)\approx\gamma_0+\gamma_0^\prime z~~~~~~~~~~~~~~0\leq
 z\leq0.5\,.
 \ee
This approximation  has been studied in Refs.~\cite{dpolarski,
Gannouji, Gannouji2}, and it was found that $\gamma_0'$ is a
quasi-constant and $\gamma_0^{\prime}\simeq -0.02$ for dark energy
models with a constant equation of state.  However, for modified
gravity models, such as some scalar-tensor models, $\gamma_0'$ is
negative and can take absolute values  larger than those in models
inside General Relativity~\cite{Gannouji}, while for the $f(R)$
model $\gamma_0'$ is also negative but its value is largely
outside the range found for dark energy models in General
Relativity~\cite{Gannouji2}. Therefore, an accurate $\gamma_0'$ at
the  low redshift could provide another characteristic
discriminative signature for these models.

In this paper, we will mainly focus on the observational
constraints on $\gamma_0$ and $\gamma_0'$ from data on the growth
factor. Firstly, we will study the linear growth index with the
form $ \gamma\approx \gamma_0+\gamma_0' z$ for the DGP model.
Then, with the best fit value $\Omega_{m,0}$ from the
observational data we will discuss the theoretical values of
$\gamma_0$ and $\gamma_0'$ and the observational constraints on
them.

\section{Growth index of DGP model}\label{sec2}
For the DGP model, in general, $G_{\rm eff}$ can be written as
 \be{eq12}
 {G_{eff}}= G_{N}\left(1+\frac{1}{3\beta}\right),
 \ee
where $ \beta=1-2r_cH\left(1+\frac{\dot{H}}{3H^2}\right)$
~\cite{r27,r29,r39,r40} and the constant $r_c$ is a scale which
sets a length beyond which gravity starts to leak out into the
bulk. According to Ref.~\cite{r40},  ${G_{\rm eff}\over G_{N} }$
can be rewritten as
 \be{eq14}
 1+\frac{1}{3\beta}=\frac{4\Omega_m^2-4\left(1-\Omega_k\right)^2
 +\alpha} {3\Omega_m^2-3\left(1-\Omega_k\right)^2+\alpha}\,,
 \ee
 where $
 \alpha\equiv 2\sqrt{1-\Omega_k}\left(3-4\Omega_k+2\Omega_m\Omega_k
 +\Omega_k^2\right)$,
 $\Omega_k\equiv
-k/(a^2H^2)$, and
 $\Omega_m\equiv 8\pi G\rho_m/(3H^2)$. Here the spatial curvature
$k=0,k>0$ and $k<0$ correspond to a flat, closed  and open
universe respectively.

For the DGP model, the modified Friedmann equation takes the
form~\cite{r33,weihao}
 \be{friedman01}
 H^2+\frac{k}{a^2}-{1\over r_c} \sqrt{H^2+{k\over a^2}}=\frac{8\pi
 G}{3}\rho_m.
 \ee
Defining $\Omega_{r_c}=\frac{1}{4r_c^2H^2_0}$, we have
 \be{eq8}
E^2(z)\equiv\left(\frac{H}{H_0}\right)^2=
\left[\sqrt{\Omega_{m,0}(1+z)^3+\Omega_{r_c}}
 +\sqrt{\Omega_{r_c}}\right]^2+\Omega_{k0}(1+z)^2\,.
 \ee
 Setting $z=0$ in the above gives rise to a constraint equation
 \be{relationoffract}
 1=\left[\sqrt{\Omega_{m,0}+\Omega_{r_c}}
 +\sqrt{\Omega_{r_c}}\right]^2+\Omega_{k0}.
 \ee
Therefore, there are only two model independent parameters out of
$\Omega_{m,0}$, $\Omega_{r_c}$ and
 $\Omega_{k0}$.

\begin{center}
 \begin{figure}[htbp]
 \centering
 \includegraphics[width=0.45\textwidth]{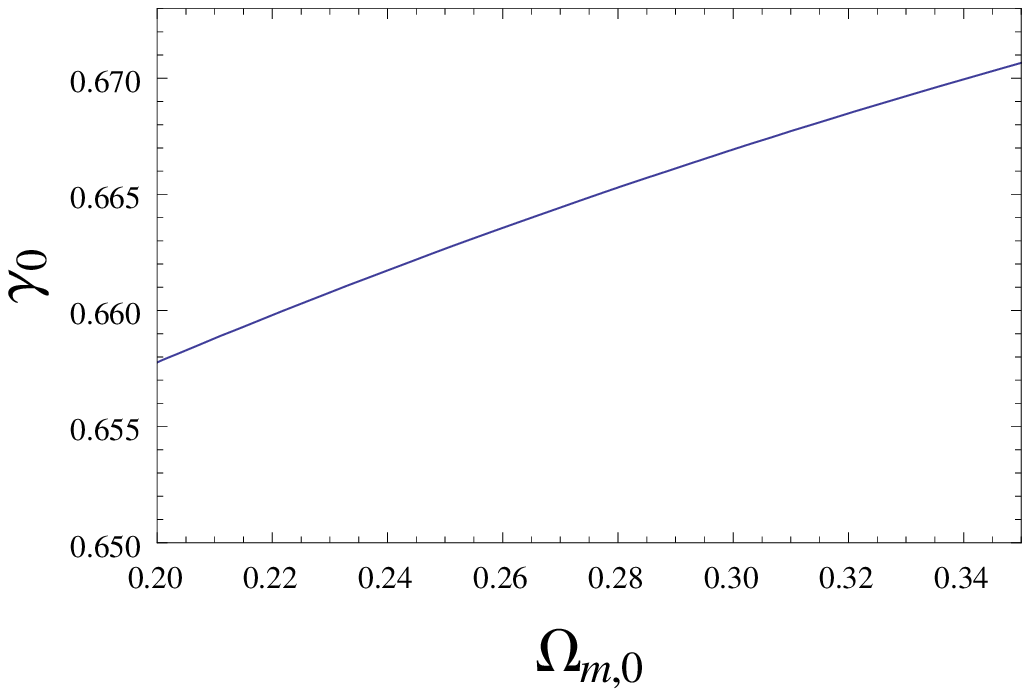}
 \includegraphics[width=0.45\textwidth]{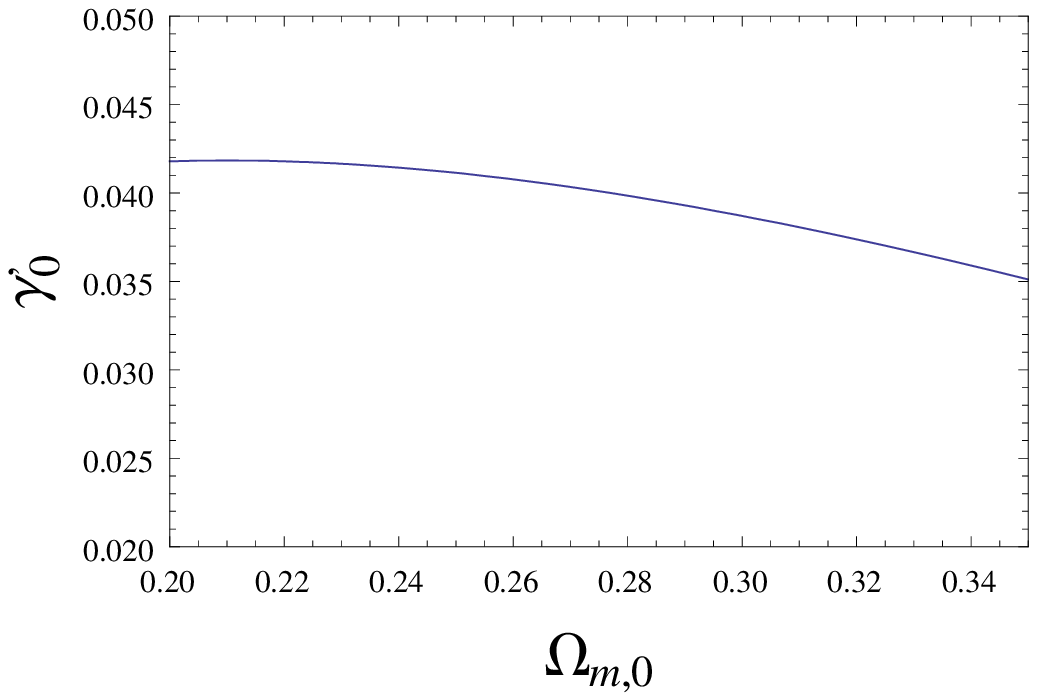}
 \caption{\label{figgamma0}  $\gamma_0$ and $\gamma_0^\prime$ are  displayed as a function of $\Omega_{m,0}$ for DGP model respectively. }
 \end{figure}
 \end{center}

 The matter density perturbation in the DGP model satisfies the
equation~\cite{wangli,r31}:
 \be{eq16}
{d\,^2\ln\delta\over d(\ln a)^2}
 +\left({d\ln\delta\over d\ln a}\right)^{ 2}
 +\left(2+\frac{d \;\ln H}{d\ln a}\right)\left(d\ln\delta\over d\ln a\right)
 =\frac{3}{2}\left(1+\frac{1}{3\beta}\right)\Omega_m\,.
 \ee
Using
 \be{eq17}
 \frac{d \;\ln H}{ d\ln a}=\frac{\dot{H}}{H^2}=-\frac{3}{2}
 +\frac{\Omega_k}{2}-\frac{3}{2} \frac{-1+\Omega_k}{1+\Omega_m-\Omega_k}\left(1-\Omega_k
 -\Omega_m\right),
 \ee
we obtain
 \begin{eqnarray}
 \label{eq18}
 &&{d^2\ln\delta\over d(\ln a)^2}
 +\left({d\ln\delta\over d\ln a}\right)^{ 2}
 +{d\ln\delta\over d\ln a}\left(\frac{1}{2}\left(1+
 \Omega_k\right)-\frac{3}{2} \frac{-1+\Omega_k}{1+\Omega_m-\Omega_k} \left(1-\Omega_k
 -\Omega_m\right)\right)\nonumber\\
 &&\quad =\frac{3}{2}\left(1+\frac{1}{3\beta}\right)\Omega_m.
 \end{eqnarray}
 Thus, according to the definition of $f$,
we have the following differential equation
\begin{eqnarray}
 \label{omegadgpdf}
 &&\Omega_m\bigg[\frac{3(-1+\Omega_k)}{1+\Omega_m-\Omega_k} (1-\Omega_k-\Omega_m)-\Omega_k\bigg]{df\over
 d\Omega_m}+f^2\nonumber\\
 &&+f\bigg[{1\over 2}(1+\Omega_k)-{3\over 2}\frac{-1+\Omega_k}{1+\Omega_m-\Omega_k}
 (1-\Omega_k-\Omega_m)\bigg]\nonumber\\
 &&={3\over2}\bigg(1+{1\over3\beta}\bigg)\Omega_m\;.
 \end{eqnarray}
Substituting  the generic expression for $f$,
Eq.~(\ref{fommegam1}), into the Eq.~(\ref{omegadgpdf}) we arive at
an equation on $\gamma (z)$
 \begin{eqnarray}
  &&{1\over
  2}[(1+\Omega_k-2\gamma\Omega_k)+
\frac{3(-1+\Omega_k)}{1+\Omega_m-\Omega_k}
(2\gamma-1)(1-\Omega_k-\Omega_m)]
\nonumber\\&&-(1+z)\gamma^\prime\ln{\Omega_m}+\Omega_m^\gamma={3\over
 2}(1+{1\over{3\beta}})\Omega_m^{1-\gamma}\,.
 \end{eqnarray}
If we only consider the linear  expansion at the low redshift as
given in Eq.~(\ref{gammaz}), it is easy to derive
 \begin{eqnarray}
 \label{gamma0prime-1}
  \gamma_0^\prime&=&(\ln\Omega_{m,0}^{-1})^{-1}\bigg[-\Omega_{m,0}^{\gamma_0}+{3\over
  2}(1+{1\over 3\beta})\Omega_{m,0}^{1-\gamma_0}-{1\over
  2}(1+\Omega_{k,0}-2\gamma_0\Omega_{k,0})\nonumber\\
  &&-3\frac{-1+\Omega_{k,0}}{1+\Omega_{m,0}-\Omega_{k,0}}
  (1-\Omega_{k,0}-\Omega_{m,0})(\gamma_0-{1 \over
  2})\bigg]\,.
 \end{eqnarray}
 This gives a constraint equation
\begin{eqnarray}
 g(\gamma_0,\gamma_0^\prime,\Omega_{m,0},\Omega_{k,0})=0\,.
 \end{eqnarray}
So, for any given background parameters $\Omega_{m,0}$ and
$\Omega_{k,0}$, the value of $\gamma_0^{\prime}$ can be determined
by  that of $\gamma_0$. For the sake of simplicity, we will only
consider the case of a  spatially  flat universe in this paper
($\Omega_k=0$). Thus from Eq.~(\ref{gamma0prime-1}),  we get
  \begin{eqnarray}
  \label{gamma0prime2}
  \gamma_0^\prime&=&(\ln\Omega_{m,0}^{-1})^{-1}\bigg[-\Omega_{m,0}^{\gamma_0}+{3\over
  2}\,{4\Omega_{m,0}^2+2\over 3\Omega_{m,0}^2+3}\,\Omega_{m,0}^{1-\gamma_0}-{1\over
  2}\nonumber\\
  &&+\,{3\over
1+\Omega_{m,0}}(1-\Omega_{m,0})(\gamma_0-{1 \over
  2})\bigg]\,.
 \end{eqnarray}
 According to equation
$f(z=0)=\Omega_{m,0}(0)^{\gamma_0}$, the value of $\gamma_0$ can be
obtained by solving Eq.~(\ref{omegadgpdf}) numerically  for an given
value of $\Omega_{m,0}$. Then plugging this obtained $\gamma_0$ into
Eq.~(\ref{gamma0prime2}), we can get the value of $\gamma_0^\prime$.
The results are shown in Fig.~\ref{figgamma0}. We find, from the
right panel, that the value of $\gamma_0$ increases from $0.658$ to
$0.671$ for $0.2\leq\Omega_{m,0}\leq0.35$. This suggests that
$\gamma$ cannot really be regarded as a constant as $\Omega_m$
varies. Notice that our result is different from that obtained for
the $\Lambda CDM$ model where the value of $\gamma_0$ is found to
decrease from 0.558 to 0.554 for
$0.2\leq\Omega_{m,0}\leq0.35$~\cite{dpolarski}. This feature of
$\gamma_0$ also provides  a distinctive signature for the DGP model
from the $\Lambda CDM$ model. From the right panel, we can see that
the $\gamma_0^\prime$ is positive and ranges approximately from
$0.035$ to $0.042$, which is also different from the dark energy
model, the scalar-tensor model and f(R) model.  For example for the
$w CDM$ model with $\Omega_{m,0}=0.3$,
 $\gamma_0^\prime$ is negative and quasi-constant
$\gamma_0^\prime\simeq-0.02$. So, in principle, we can
discriminate the DGP model from the dark energy model merely
through the sign of $\gamma_0^\prime$ if we can have an accurate
value of $\gamma_0^\prime$ from the observation data. Now we will
discuss the the observational constraints on $\gamma_0$ and
$\gamma'_0$

%============================= Fig. 2 =================================

 \section{Observational constraints}
In order to obtain the observational constraints on $\gamma_0$ and
$\gamma'_0$, we firstly need to know the value of $\Omega_{m,0}$
determined by  observations. Here we use the results  in
Ref.~\cite{gongyungui} where the author found
$\Omega_{m,0}=0.273\pm 0.015$ for the $\Lambda CDM$ model and
$\Omega_{m,0}=0.278\pm 0.015$ for the DGP model respectively  from
the 307 Union Sne Ia data, the BAO from the SDSS data, the shift
parameter from the WMAP5 and the 11 Hubble parameter data. Using
the best fit values $\Omega_{m,0}=0.273$ for the $\Lambda$CDM
model and $\Omega_{m,0}=0.278$ for the DGP model respectively,  we
find $\gamma_0=0.665, \gamma_0^\prime=0.04$ for theoretical vaules
for the DGP model from  Fig. (1) in this paper, and
$\gamma_0=0.555, \gamma_0^\prime=-0.018$ for the $\Lambda CDM$
model from  Fig. (1) in Ref.~\cite{dpolarski}.

To find the observational constraints on $\gamma_0$ and
$\gamma_0'$, only three observational data on $f_{obs}$ given in
Table I can be used, since the linear expansion is valid only at
the low reshifts.  With the best fit value of $\Omega_{m,0}$ we
can obtain the constraints from the observations by using the
following equation
 \begin{equation}
\label{fzchi}
\chi^2_f=\sum_{i=1}^{3}\frac{[f_{obs}(z_i)-\Omega_m^{\gamma_0+\gamma_0^\prime
z_i}]^2}{\sigma_{fi}^2},
\end{equation}
 where $\sigma_{fi}$
is the $1 \sigma$ uncertainty of the $f(z)$ data. The results are
shown in Fig. (2). The best fit values are $\gamma_0=0.774,
\gamma_0'=-0.556$ for the $\Lambda CDM$ model and $\gamma_0=0.767,
\gamma_0'=-0.732$ for the DGP model, which show that the
observations imply an negative value of $\gamma_0'$. Since the DGP
model gives an positive  $\gamma_0'$, thus we can conclude that
observations disfavor the DGP model. However, from Fig. (2), we
find that at the $1\sigma$ confidence level both the $\Lambda CDM$
and the DGP model are consistent with the observations.
\begin{table}[htp]
\begin{tabular}{|c|c|c|}
\hline
\ \ \ \ \ \ \ \ \   $z$\ \ \ \ \ \ \ \ \   & \ \ \ \ \ \ \  $f_{obs}$\ \ \ \ \ \ \  &\ \ References \\
\hline
$0.15$ & $0.49\pm 0.1$ & \cite{guzzo} \\
\hline
$0.35$ & $0.7\pm 0.18$ & \cite{tegmark} \\
\hline
$0.55$ & $0.75\pm 0.18$ & \cite{ross} \\
\hline
\end{tabular}
\caption{The summary of the observational data on the growth factor
$f$ at low redshifts.} \label{fzdata}
\end{table}

\begin{center}
 \begin{figure}[htbp]
 \centering
 \includegraphics[width=0.45\textwidth]{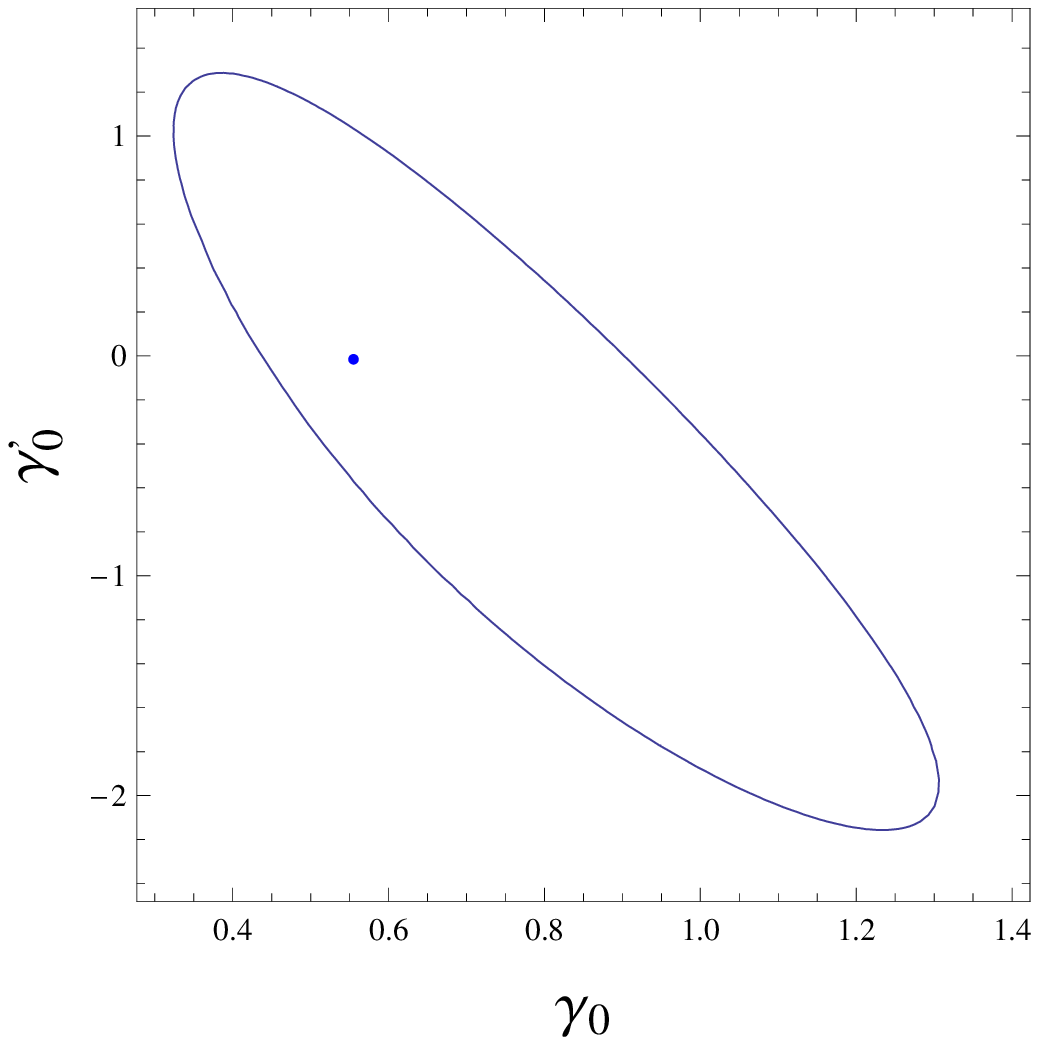}
 \includegraphics[width=0.45\textwidth]{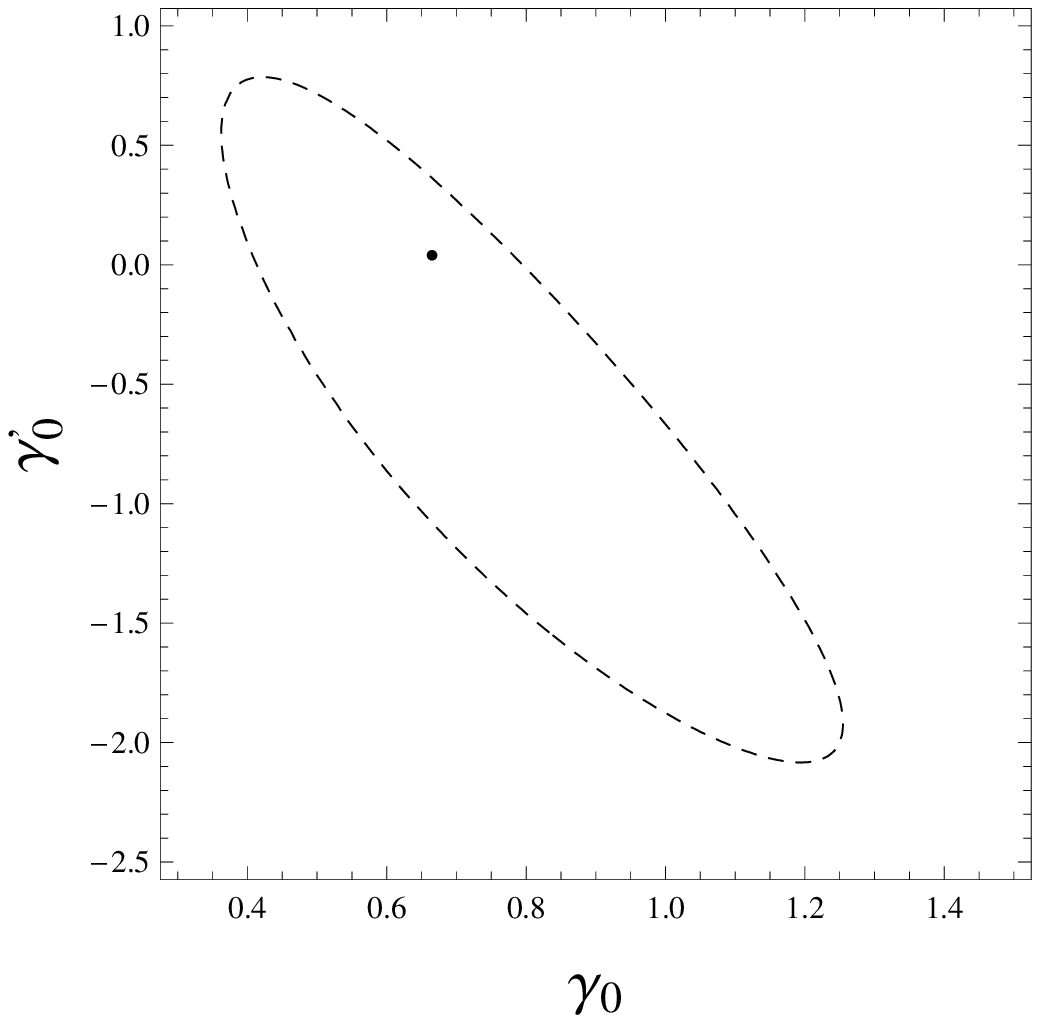}
 \caption{\label{constraints}  The $1 \sigma$ contours of $\gamma_0$ and
  $\gamma_0^\prime$ by fitting the $\Lambda$CDM model and DGP model  to the growth rate data.
 The points denote  the theoretical values of $\gamma_0$ and
  $\gamma_0^\prime$ with the $\Omega_{m,0}$ taking the best fit values.}
 \end{figure}
 \end{center}

\section{Conclusion}\label{sec4}
In this letter, the growth factor of matter perturbations in the DGP
model is studied and we find that the growth index, $\gamma$, should
be treated as a function of time. With a linear expansion of
$\gamma(z)\approx \gamma_0+\gamma_0^\prime z$, we obtain that
$\gamma_0$ increases from $0.658$ to $0.671$ and $\gamma_0^\prime$
ranges approximately from $0.035$ to $0.042$, for
$0.2\leq\Omega_{m,0}\leq0.35$.  This is different from the results
obtained for the $\Lambda CDM$ model where $\gamma_0$  decreases
from 0.558 to 0.554 and $\gamma_0^\prime$ is quasi-constant with
$\gamma_0^\prime\simeq-0.02$ for
$0.2\leq\Omega_{m,0}\leq0.35$~\cite{dpolarski}. These features
provide distinctive signatures for the DGP model from the $\Lambda
CDM$ model. With the observational data on the growth factor, we
analyze the observational constraints on $\gamma_0$ and $\gamma_0'$
and find that the best fit values are $\gamma_0=0.774,
\gamma_0'=-0.556$ for the $\Lambda CDM$ model and $\gamma_0=0.767,
\gamma_0'=-0.732$ for the DGP model. This seems to show that the
observations favor the $\Lambda CDM$ model since the theoretical
value of $\gamma_0'$ is positive for the DGP model. However, at
$1\sigma$ confidence level both the $DGP$ model and $\Lambda CDM$
model are consistent with the observations as can be seen from
Fig~{\ref{constraints}}. It should be pointed out that our results
are based upon merely three low redshifts data, since the linear
approximation is valid only at the low redshiftes. To obtain
stronger constraints which can clearly discriminate different
models, we need a parametrized form of $\gamma(z)$, which is
applicable for all the observational data, and hope to turn to this
issue in the future~\cite{FYW}.

\section*{Acknowledgments}
X. Fu is grateful to Professor Yungui Gong for his very helpful
discussions. This work was supported in part by the National
Natural Science Foundation of China under Grants No. 10575035,
10775050, 10705055, the SRFDP under Grant No. 20070542002, the
Research Fund of Hunan Provincial Education Department,  the Hunan
Provincial Natural Science Foundation of China under Grant No.
08JJ4001, and the China Postdoctoral Science Foundation.

\end{document}